\documentclass[aps,preprint,tightenlines,nofootinbib,amssymb,superscriptaddress]{revtex4-1}
\usepackage{amssymb}
\usepackage{mathrsfs}
\usepackage{slashed}
\usepackage{graphicx,color}
\usepackage{amsmath}
\usepackage[colorlinks=true,linkcolor=blue,citecolor=blue]{hyperref}%
\allowdisplaybreaks

\begin{document}
\title{Lepton flavor violation induced by neutral and
doubly-charged scalars at future lepton colliders}

\author{P. S. Bhupal Dev}
\affiliation{Department of Physics and McDonnell Center for the Space Sciences,  Washington University, St. Louis, MO 63130, USA}

\author{Rabindra N. Mohapatra}
\affiliation{Maryland Center for Fundamental Physics, Department of Physics, University of Maryland, College Park, MD 20742, USA}

\author{Yongchao Zhang\footnote{Talk presented at the International Workshop on Future Linear Colliders (LCWS2018), Arlington, Texas, 22-26 October 2018. C18-10-22.}}
\affiliation{Department of Physics and McDonnell Center for the Space Sciences,  Washington University, St. Louis, MO 63130, USA}
\affiliation{Center for High Energy Physics, Peking University, Beijing 100871, China}

\date{\today}

\begin{abstract}
  New physics scenarios beyond the Standard Model (SM) for neutrino mass mechanism often necessitate the existence of a neutral scalar $H$ and/or doubly-charged scalar $H^{\pm\pm}$, which couple to the SM charged leptons in a flavor violating way, while evading all existing constraints. Such scalars could be effectively produced at future lepton colliders like CEPC, ILC, FCC-ee and CLIC, either on-shell or off-shell, and induce striking charged lepton flavor violating (LFV) signals. We find that a large parameter space of the scalar masses and the LFV couplings can be probed at lepton colliders, well beyond the current low-energy constraints in the lepton sector. The neutral scalar explanation of the muon $g-2$ anomaly could also be directly tested.
  \end{abstract}
\maketitle


\section{Introduction}

There are various theoretical models of new physics which lead to charged lepton flavor violation (cLFV) effects at an observable level, which is strongly implied by the observation of flavor violation in the neutral lepton sector, i.e.~the solar, atmospheric and reactor neutrino oscillations~\cite{PDG}.  In the minimal extension of Standard Model (SM) with Dirac neutrinos, cLFV rates are highly suppressed due to the tiny neutrino masses. This makes the experimental searches for  cLFV all the more interesting, because any observable effect must come from physics beyond the minimally extended SM related to the origin of neutrino mass. LFV often arises from the extension of the Higgs sector, which allows flavor-violating Yukawa couplings of new neutral and/or doubly-charged scalars beyond the SM. In this report, we summarize the main results of {\it model-independent} LFV prospects at future lepton colliders, such as the Circular Electron-Positron Collider (CEPC)~\cite{CEPC-SPPCStudyGroup:2015csa}, International Linear Collider (ILC)~\cite{Baer:2013cma}, Future Circular Collider (FCC-ee)~\cite{Gomez-Ceballos:2013zzn} and Compact Linear Collider (CLIC)~\cite{Battaglia:2004mw}, based on our recent papers~\cite{Dev:2017ftk, Dev:2018upe}. This complements previous studies of LFV at lepton colliders that are performed in the framework of effective four-fermion couplings~\cite{Kabachenko:1997aw, Cho:2016zqo, Ferreira:2006dg, Aranda:2009kz, Murakami:2014tna} or in the context of flavor-violating SM Higgs decays~\cite{Banerjee:2016foh, Chakraborty:2016gff, Chakraborty:2017tyb, Qin:2017aju} and tau decays~\cite{Hays:2017ekz} or with doubly-charged scalars~\cite{Rodejohann:2010bv, Nomura:2017abh}. Compared to the hadron colliders, the lepton colliders are generally very ``clean'' and the SM processes therein are well understood, which render them primary facilities to search for new physics  via the LFV signals.

If any of the new neutral scalars $H$ is hadrophobic (i.e. couples dominantly to leptons), it could remain sufficiently light and contribute sizably to cLFV with the Yukawa couplings $h_{\alpha\beta}$ ($\alpha,\,\beta = e,\,\mu,\,\tau$ being the lepton flavor indices), while easily evading the direct searches at hadron colliders, as well as the low-energy quark flavor constraints, such as the rare flavor-changing decays and oscillations of $K$ and $B$ mesons.
Some well-motivated examples include supersymmetric models with leptonic $R$-parity violation~\cite{susy, susy2, susy3, susy4}, left-right symmetric models~\cite{Dev:2016dja, Dev:2016vle, Dev:2017dui, Maiezza:2016ybz}, two-loop models for neutrino masses~\cite{babu}, mirror models~\cite{mirror, mirror2, mirror3, mirror4, mirror5}, and two-Higgs doublet models~\cite{2HDM, 2HDM2}, where the cLFV coupling might arise at tree and/or loop level.

The doubly-charged scalar and its LFV Yukawa couplings $f_{\alpha\beta}$ might be closely related to neutrino mass generation, e.g. in the type-II seesaw~\cite{type2a,type2b,type2c,type2d,type2e} and its left-right extension~\cite{LR, Mohapatra:1974gc, Senjanovic:1975rk}. The magnitudes of $f_{\alpha\beta}$ might be sizable and hence accessible at lepton colliders.  There are of course constraints on some of these couplings from rare lepton decays like $\ell_\alpha \to \ell_\beta \gamma$ and $\ell_\alpha \to \ell_\beta \ell_\gamma \ell_\delta$, but they leave enough room for some of them being of order one. Furthermore, rare lepton decays generally probe products of two different $f$ couplings whereas the lepton collider probes them individually~\cite{Dev:2017ftk, Dev:2018upe}. We show the interesting ranges of effective Yukawa couplings $h_{\alpha\beta}$ and $f_{\alpha\beta}$ that can be measured in the planned lepton colliders such as CEPC and ILC and can provide new ways to test the underlying extended Higgs models. They will in any case provide complementary information to rare lepton decay constraints on the $h$ and $f$ couplings, which makes such studies interesting from the synergistic viewpoint of energy and intensity frontiers.

This paper is organized as follows: Section~\ref{sec:neutral} is devoted to the beyond SM neutral scalar $H$, where we consider  the LFV from both the on-shell and off-shell production of $H$ at lepton colliders. As two benchmark setups for the future lepton colliders, we show the corresponding prospects at CEPC 240 GeV and ILC 1 TeV, with  total integrated luminosity of 5 ab$^{-1}$ and 1 ab$^{-1}$, respectively. In Section~\ref{sec:dcs}, we focus on the  single, pair and off-shell production of doubly-charged scalar $H^{\pm\pm}$ at lepton colliders through the Yukawa couplings $f_{\alpha\beta}$ and the resultant LFV prospects at ILC and CEPC. We conclude in Section~\ref{sec:conclusion}. More details can be found in Refs.~\cite{Dev:2017ftk, Dev:2018upe}.

\section{LFV induced by new neutral scalar}
\label{sec:neutral}

Without loss of generality, we can write the effective Yukawa couplings of the neutral scalar $H$ to the charged leptons as
\begin{eqnarray}
\label{eqn:Yukawa}
{\cal L}_Y \ = \ h_{\alpha \beta}  \bar{\ell}_{\alpha,\,L} H \ell_{\beta,\, R} ~+~ {\rm H.c.} \, .
\end{eqnarray}
Here for simplicity we assume the couplings $h_{\alpha\beta}$ are all real and chirality-independent and thus symmetric. The scalar $H$ may or may not be responsible for symmetry breaking and/or mass generation of other particles in realistic models, where it could be part of a singlet, doublet or triplet scalar field.  We assume further that $H$ is CP-even and its mixing with and coupling to the SM Higgs is small.  If the scalar is CP-odd, the limits and prospects derived here would not change significantly. Even though there are all varieties of stringent low-energy cLFV constraints, such as $\ell_\alpha\to \ell_\beta \gamma,~ 3\ell_\beta, ~\ell_\beta\ell_\beta\ell_\gamma$, only a few of them are directly relevant to the LFV prospects discussed below.
With an ab$^{-1}$ level of integrated luminosity at future lepton colliders,  a large parameter space of the scalar mass $m_{H}$ and $h_{\alpha\beta}$ could be probed, well beyond the current cLFV constraints, and complementary to the projected low-energy constraints from future experiments at the intensity frontier~\cite{Calibbi:2017uvl}.
In addition, the Lagrangian in Eq.~(\ref{eqn:Yukawa}) also gives rise to a one-loop contribution to the lepton anomalous magnetic moment. In particular, the scalar field coupling whose  induced loop graph helps resolve the longstanding muon $g-2$ discrepancy can be directly tested at lepton colliders.

\subsection{On-shell LFV}
If kinematically allowed, the neutral scalar $H$ can be directly produced at lepton colliders, in association with a pair of flavor-changing leptons through the couplings in Eq.~(\ref{eqn:Yukawa}), i.e.
\begin{eqnarray}
\label{eqn:neutral1}
e^+ e^- \to \ell^\pm_\alpha \ell^\mp_\beta + H
\end{eqnarray}
mediated by the SM photon or $Z$ boson,\footnote{We have also the production channel $e^+ e^- \to (\gamma/Z) + H$~\cite{Dev:2018upe}. However, this process involves only the flavor-conserving coupling $h_{ee}$, and we will not consider it here.} and
\begin{eqnarray}
\label{eqn:neutral2}
e^+ e^- \to \nu_\alpha \bar{\nu}_\beta + H \quad \text{(with $\alpha = e$ or $\beta = e$)}
\end{eqnarray}
mediated by the SM $W$ boson. In future lepton colliders, high luminosity photon beams can also be obtained by Compton back-scattering of low-energy, high-intensity laser beam off the high-energy electron beam~\cite{Ginzburg:1981vm, Ginzburg:1982yr, Telnov:1989sd}, and the neutral scalar $H$ can also be produced through the processes
\begin{eqnarray}
\label{eqn:neutral3}
e^\pm \gamma \to \ell_\alpha^\pm + H \quad \text{and} \quad \gamma\gamma \to \ell_\alpha^\pm \ell_\beta^\mp + H \,.
\end{eqnarray}

\begin{figure*}[!t]
  \centering
  \includegraphics[width=0.4\textwidth]{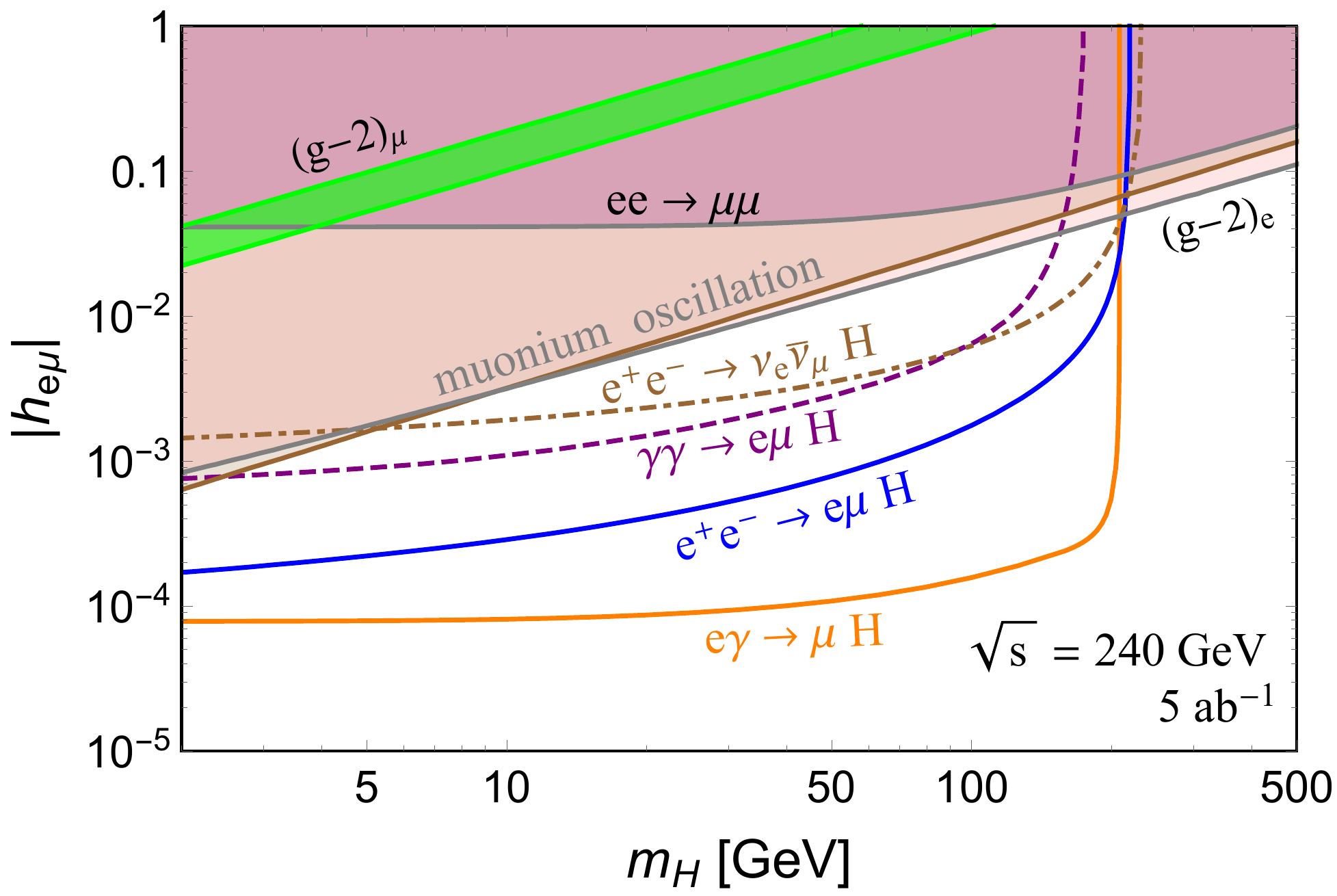}
  \includegraphics[width=0.4\textwidth]{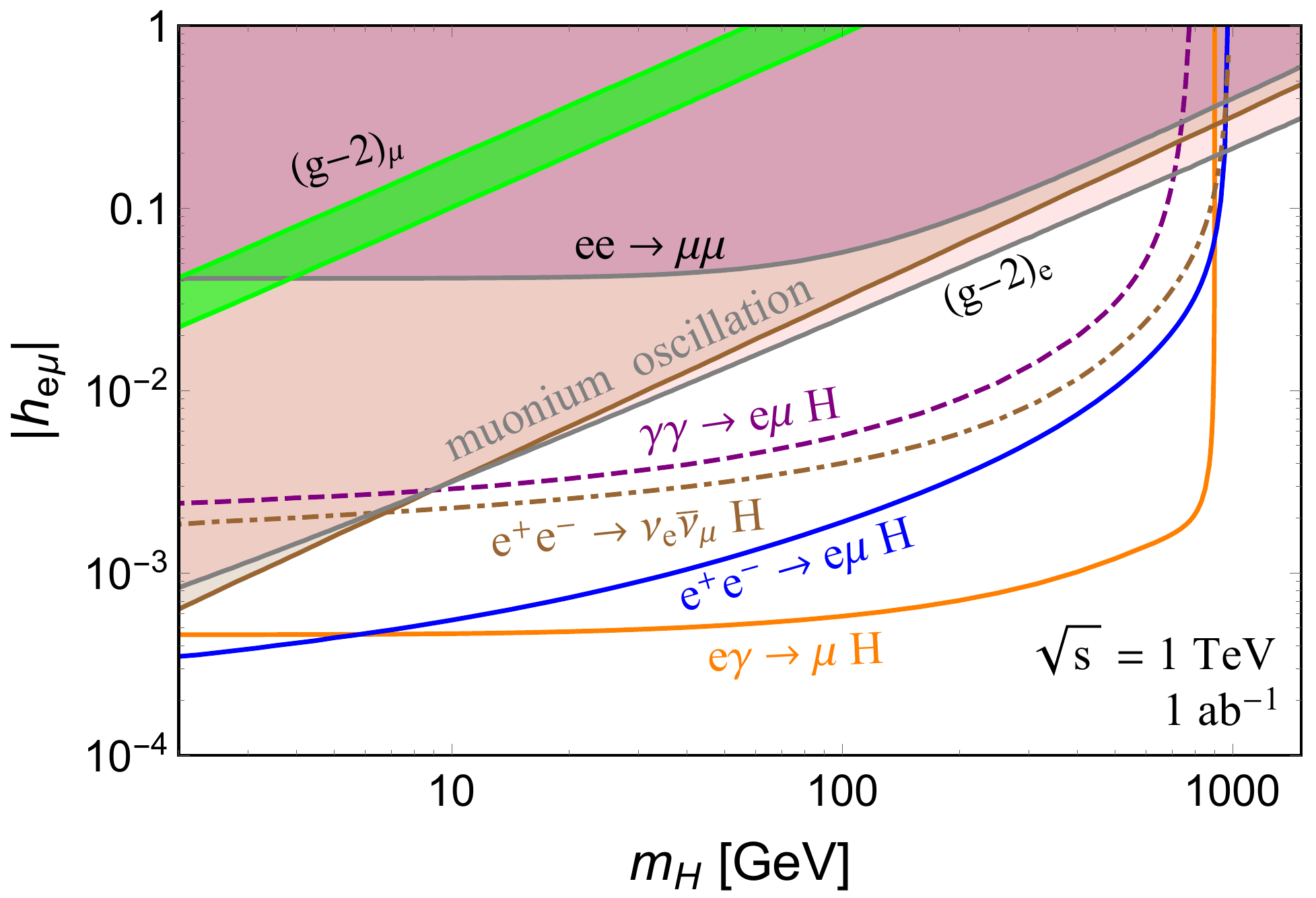} \vspace{-2pt} \\
  \includegraphics[width=0.4\textwidth]{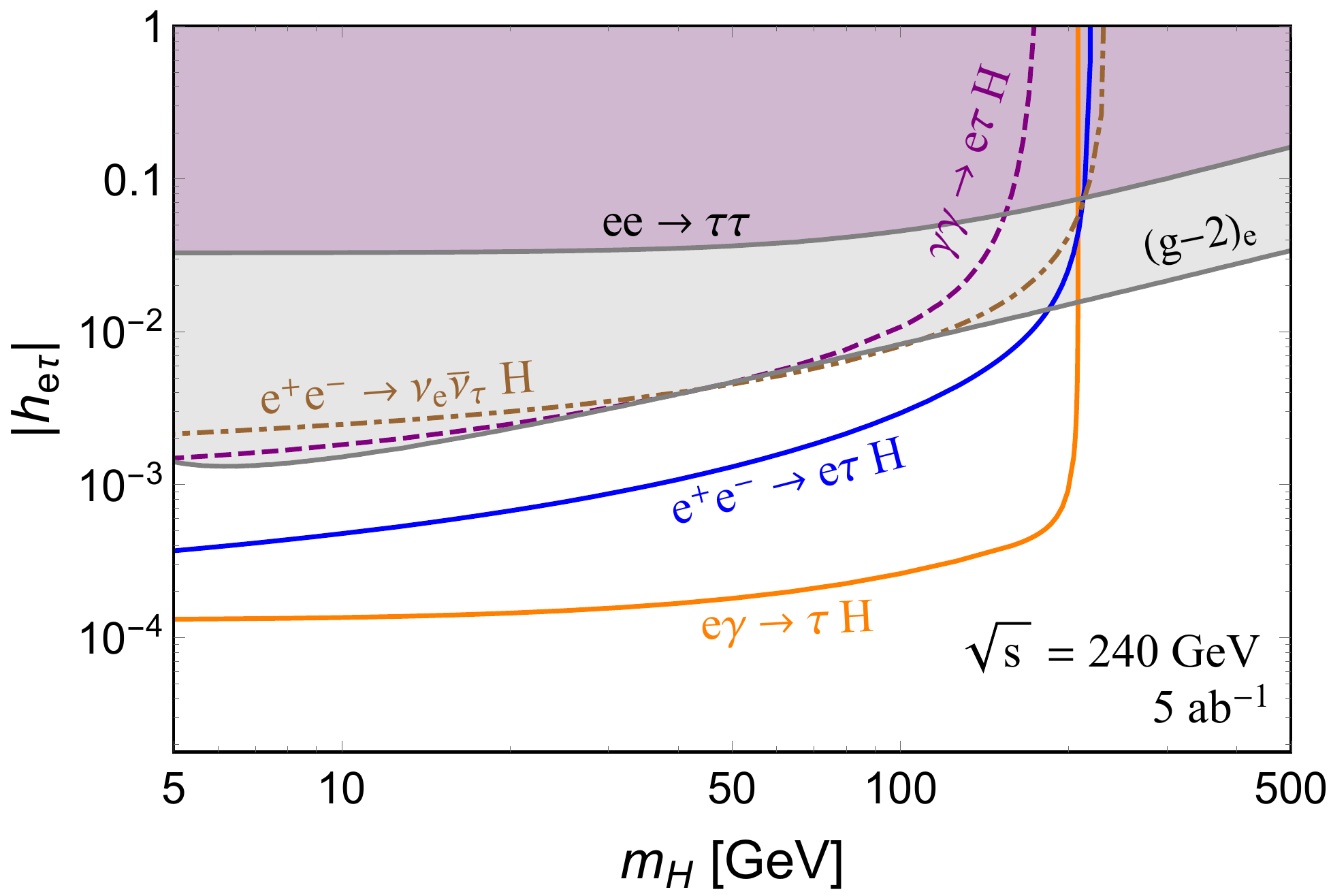}
  \includegraphics[width=0.4\textwidth]{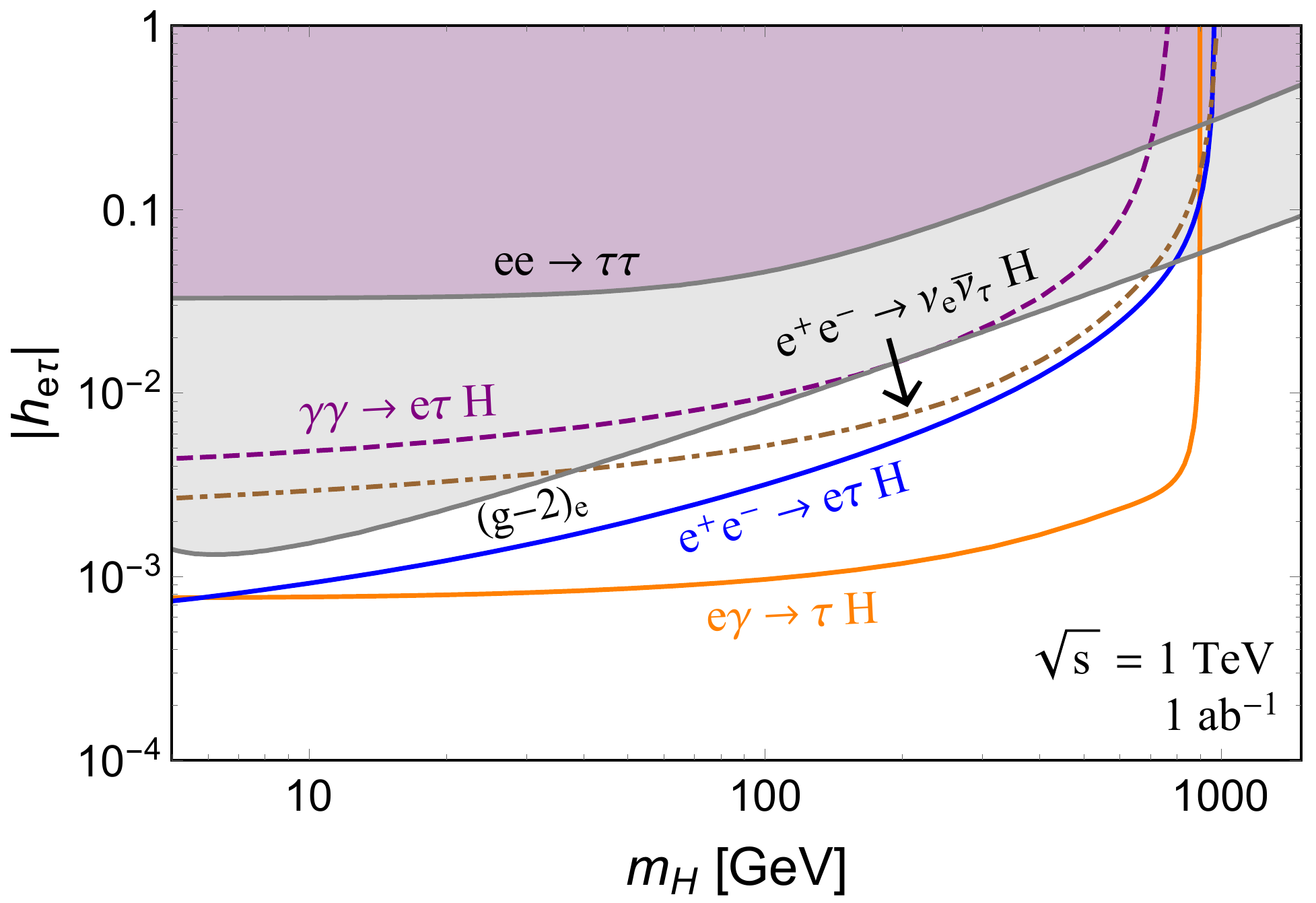} \vspace{-2pt} \\
  \includegraphics[width=0.4\textwidth]{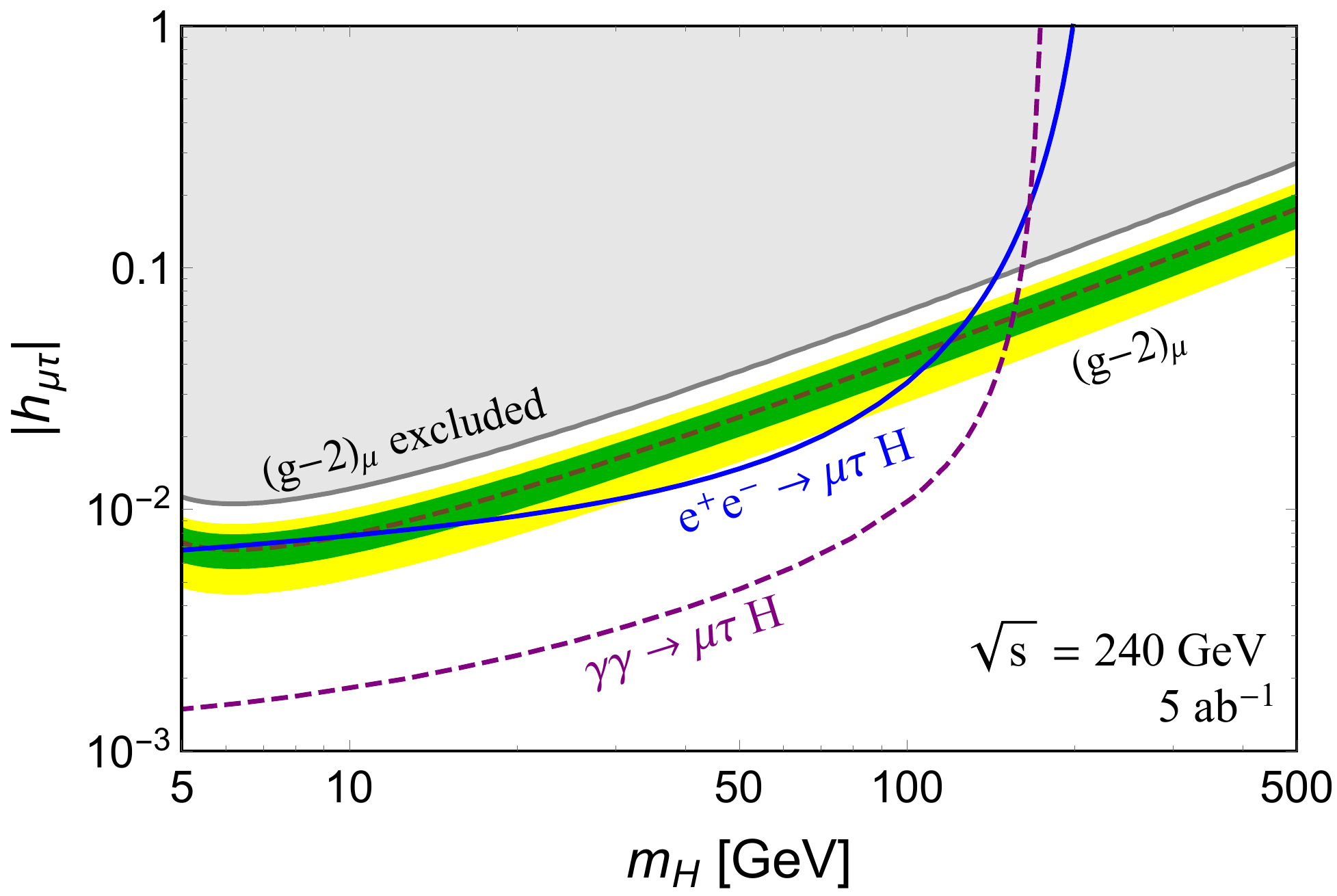}
  \includegraphics[width=0.4\textwidth]{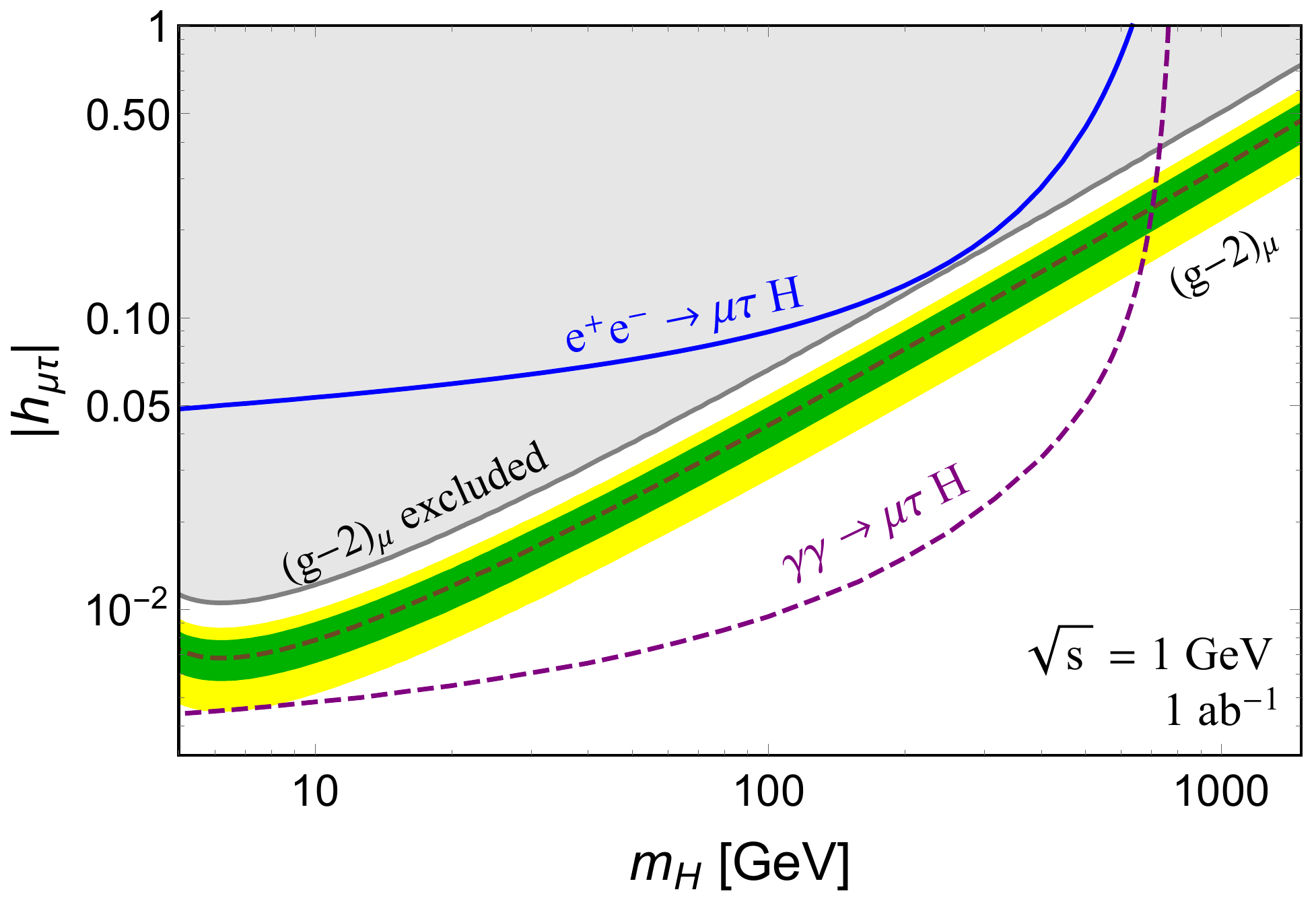}
  \caption{Prospects of the couplings $h_{\alpha\beta}$ from on-shell production of $H$ at CEPC (240 GeV and 5 ab$^{-1}$, left) and ILC (1 TeV and 1 ab$^{-1}$, right), in the channels of $e\gamma \to \ell H$, $e^+ e^-,\, \gamma\gamma \to \ell_\alpha^\pm \ell_\beta^\mp H$ and $e^+ e^- \to \nu \bar\nu H$. The shaded regions are excluded by the muonium oscillation, electron $g-2$, muon $g-2$ (excluded by the theoretical-experimental discrepancy at the $5\sigma$ CL)~\cite{PDG} and the LEP $e^+e^- \to \ell^+ \ell^-$ data~\cite{Abdallah:2005ph}. The green and yellow bands in the first and third rows can explain the muon $g-2$ anomaly at the $1\sigma$ and $2\sigma$ CL, respectively, while the dotted line at the center of the $1\sigma$ band corresponds to the central value. }
  \label{fig:p1}
\end{figure*}

Let us first consider the coupling $h_{e\mu}$, switching off all other (LFV) couplings. It should be emphasized that all the production amplitudes in Eqs.~(\ref{eqn:neutral1}) to (\ref{eqn:neutral3}) depend only on the LFV couplings $h_{\alpha\beta}$ (here $\alpha\beta = e\mu$), and thus could be easily made to satisfy the rare lepton decay constraints, such as those from $\mu \to eee$ and $\mu \to e \gamma$, which depend on the product $|h_{ee}^\dagger h_{e\mu}|$. With vanishing or suppressed couplings to the quark sector, the $\mu - e$ conversion limits are irrelevant. Furthermore, for real Yukawa couplings, we do not either have any limits from electric dipole moment. Only the following constraints are relevant:
\begin{enumerate}
\item [({\it i})] {\it Muonium-antimuonium oscillation:} This could occur in both $s$ and $t$-channels, with the oscillation probability ${\cal P} \propto |h_{e\mu}|^4/m_H^4$, and is constrained by the MACS experiment~\cite{Willmann:1998gd}.

\item [({\it ii})] $(g-2)_e$: The anomalous magnetic moment of electron $a_e$ receives a contribution from the $H - \mu$ loop, which is strongly constrained from one-electron quantum cyclotron experiments~\cite{Hanneke:2008tm}.

\item [({\it iii})] $e^+e^-\to \mu^+\mu^-$: A $t$-channel $H$ could mediate the scattering $e^+ e^- \to \mu^+ \mu^-$, which interferes with the SM diagrams in the $s$-channel and is constrained from LEP $e^+e^-\to \mu^+\mu^-$ data~\cite{Abdallah:2005ph}.
\end{enumerate}
All these constraints on $m_H$ and $h_{e\mu}$ are shown as the shaded regions in the upper two panels of Fig.~\ref{fig:p1} (which correspond to two benchmark configurations of future lepton colliders). To explain the longstanding theoretical and experimental discrepancy of the muon $g-2$, i.e. $\Delta a_\mu = (2.87 \pm 0.80) \times 10^{-9}$~\cite{PDG}, the LFV coupling $h_{e\mu}$ is required to be larger, as shown by the green band in the upper panels of Fig.~\ref{fig:p1}, which is already excluded by the electron $g-2$ and muonium oscillation constraints.

As for the future lepton collider sensitivity, we specifically consider two benchmark configurations, namely, the CEPC 240 GeV with an integrated luminosity of 5 ab$^{-1}$ and ILC 1 TeV with a luminosity of 1 ab$^{-1}$.
The SM background is dominated by particle mis-identification, e.g. electron mis-identified as muon or vice versa, and is very small (see e.g. Refs.~\cite{bkg1,bkg2,bkg3}). After being produced, $H$ could decay back into the charged lepton pairs or other SM particles. To be concrete, we consider the most optimistic case, i.e.  the neutral scalar $H$ decays predominantly into a pair of leptons, i.e. $H \to \ell_\alpha^\pm \ell_\beta^\mp$ (here $H$ decays only into $e^\pm \mu^\mp$ pair). We apply  nominal cuts $p_T (\ell) > 10$ GeV, $|\eta(\ell)| < 2.5$ and $\Delta R_{\ell\ell^\prime} > 0.4$ using {\tt CalcHEP}~\cite{Belyaev:2012qa}. The corresponding prospects of $m_H$ and $|h_{e\mu}|$ in the channels of Eqs.~(\ref{eqn:neutral1}) to (\ref{eqn:neutral3}) at CEPC and ILC are shown respectively in the upper left and upper right panels of Fig.~\ref{fig:p1}, where we have assumed a minimum of 10 signal events at both  colliders. It is clear from Fig.~\ref{fig:p1} that a large region of $m_{H}$ and $|h_{e\mu}|$ can be probed in future lepton colliders, which significantly extends the current limits.  Reconstructing the $H$ peak from the decay products could further improve the significance of the LFV signals, which however will be rather model-dependent.

Turning now to the coupling $h_{e\tau}$, the most stringent limits come from the electron $g-2$~\cite{Hanneke:2008tm}, and the LEP $e^+e^- \to \tau^+\tau^-$ data~\cite{Abdallah:2005ph}. The reconstruction of $\tau$ lepton is more challenging than $\mu$,\footnote{Following Ref.~\cite{Baer:2013cma}, we have adopted an efficiency factor of $60\%$ for the tau lepton reconstruction.} but there is still ample parameter space to probe at both CEPC and ILC, as shown in the middle panels of Fig.~\ref{fig:p1}, where we have assumed $H \to e^\pm \tau^\mp$ is the predominant decay mode.

Regarding the coupling $h_{\mu\tau}$, there are currently no experimental limits, except for the muon $g-2$ discrepancy~\cite{PDG}. This could be explained in presence of $H$ when it couples to muon and tau, as shown by the brown line (for the central value) and the green and yellow bands for the $1\sigma$ and $2\sigma$ confidence level (CL), respectively, in the lower panels of Fig.~\ref{fig:p1}, while the shaded region is excluded by the current muon $g-2$ data at the 5$\sigma$ CL. It is clear in Fig.~\ref{fig:p1} that the $(g-2)_\mu$ anomaly can be directly tested at CEPC (ILC) up to a scalar mass of $\simeq 150 \, (700)$ GeV, in particular in the $\gamma\gamma$ channel. With a larger luminosity being planned~\cite{Gomez-Ceballos:2013zzn} at FCC-ee and a higher energy at CLIC, these prospects could be even better.

\begin{figure}[!t]
  \centering
  \includegraphics[width=0.4\textwidth]{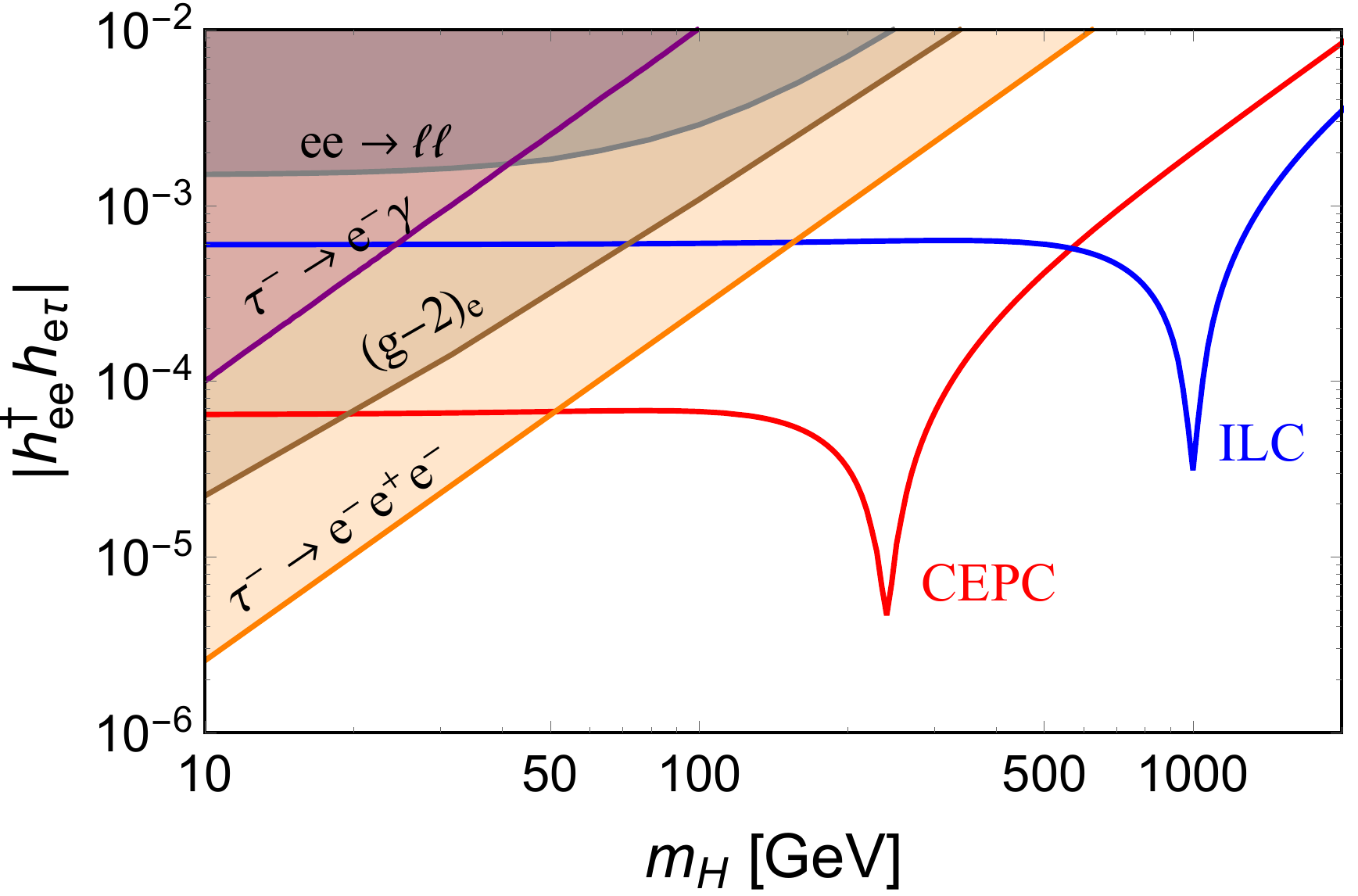}\\
  \includegraphics[width=0.4\textwidth]{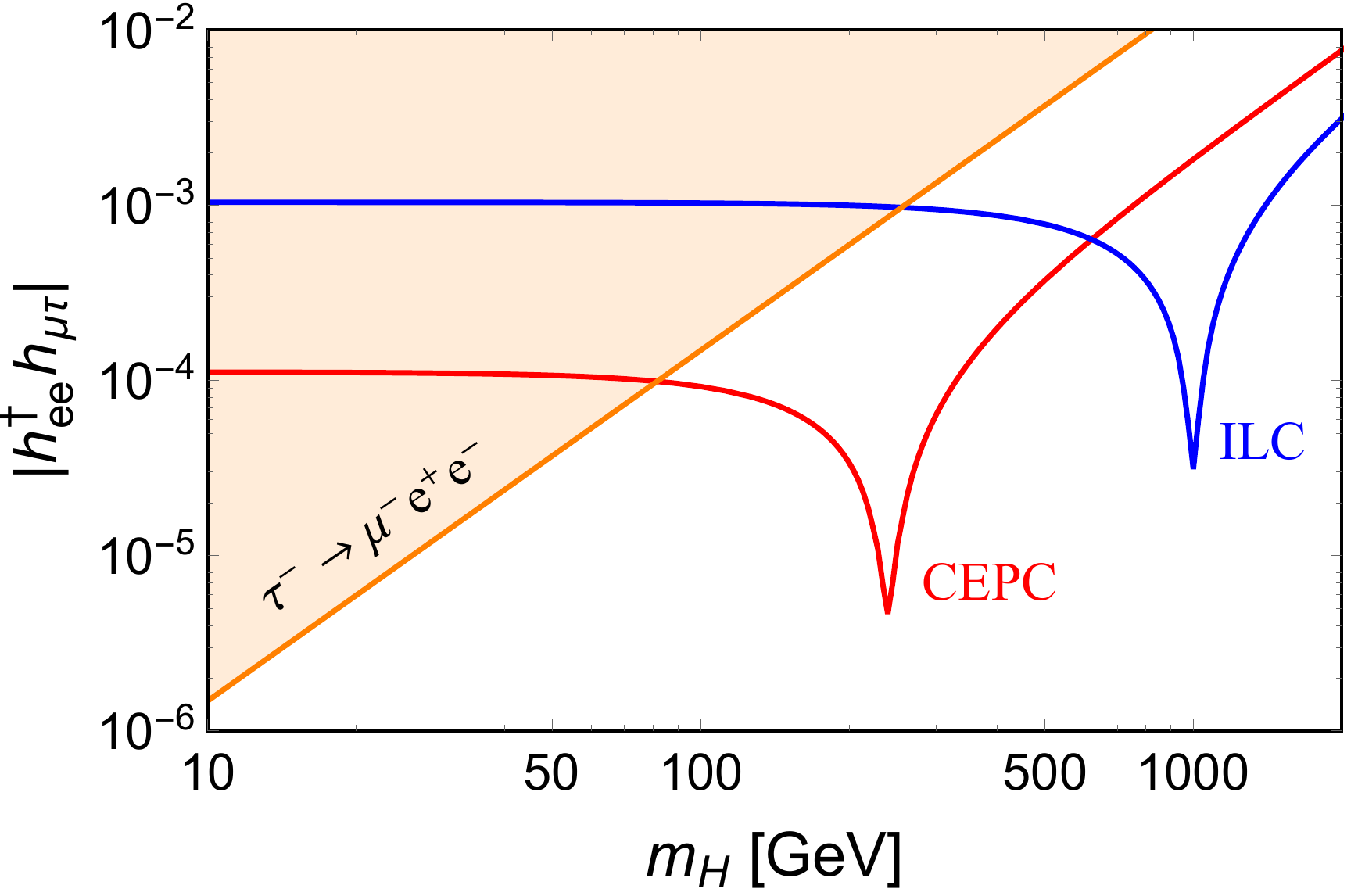}
  \includegraphics[width=0.4\textwidth]{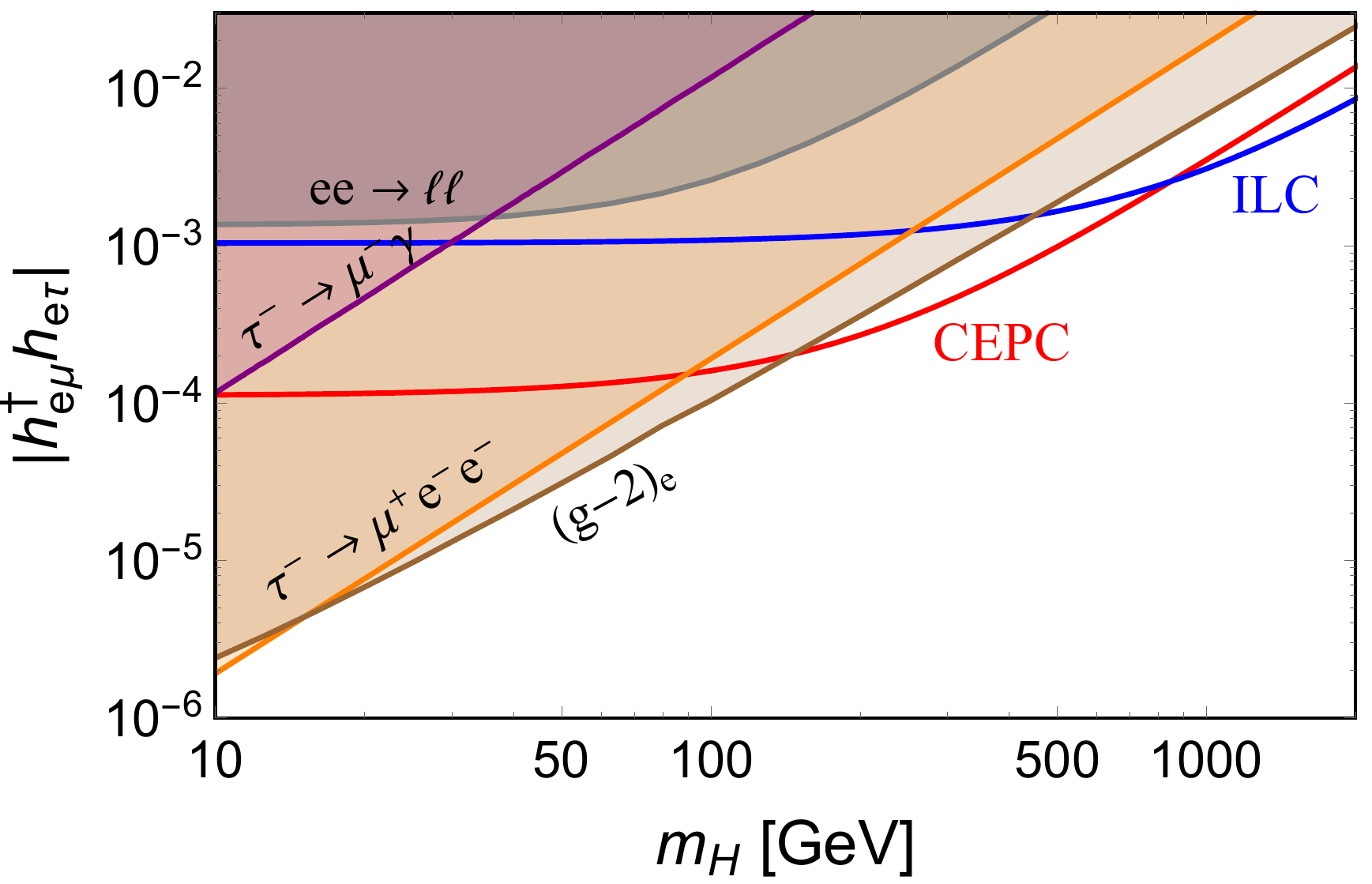}
  \vspace{-5pt}
  \caption{Prospects of $|h^\dagger_{ee} h_{e\tau}|$ (top), $|h^\dagger_{ee} h_{\mu\tau}|$ (bottom, left) and $|h^\dagger_{e\mu} h_{e\tau}|$ (bottom, right) from searches of $e^+ e^- \to e^\pm \tau^\mp,\, \mu^\pm \tau^\mp$ at CEPC 240 GeV (red) and ILC 1 TeV (blue). Here we have assumed 10 signal events. Also shown are the constraints from rare lepton decays, $(g-2)_e$, and the LEP $e^+ e^- \to \ell^+ \ell^-$ data. }
  \vspace{-5pt}
  \label{fig:p2}
\end{figure}

\subsection{Off-shell LFV}
The LFV signals could also be produced from an off-shell $H$, i.e.
$e^+ e^- \to \ell^\pm_\alpha \ell^\mp_\beta$. This could occur in both $s$ and $t$ channels, whereas in the $s$-channel, $H$ is on-shell if the colliding energy $\sqrt{s} \simeq m_{H}$, which corresponds to the resonant production $e^+ e^- \to H$ with the subsequent decay of $H \to \ell_\alpha^\pm \ell_\beta^\mp$. Different from the on-shell case, the off-shell production amplitudes have a quadratic dependence on the Yukawa couplings (some of them might be flavor conserving), and thus largely complementary to the on-shell LFV  searches.

The amplitude $e^+ e^- \to e^\pm \mu^\mp$ is proportional to $h^\dagger_{ee} h_{e\mu}$. This is tightly constrained by the $\mu \to eee$ data~\cite{PDG}, leaving no hope to see any collider LFV signal in this channel. In the tau sector, the LFV decay constraints are comparatively weaker. The limits on $|h^\dagger h|/m_{H}^2$ from $\tau^- \to e^- \gamma$ and $\tau^- \to e^+ e^- e^-$ are almost constants, as in effective field theories with superheavy mediators. As for the on-shell case above, the coupling $|h^\dagger_{ee} h_{e\tau}|$ are constrained by electron $g-2$ and the LEP $e^+ e^- \to \ell^+ \ell^-$ data~\cite{Abdallah:2005ph}. All the constraints are presented in the top panel of Fig.~\ref{fig:p2}, with the shaded regions excluded.
Note that the muon $g-2$ can not be used to set unambiguous limits on the combinations $|h_{ee}^\dagger h_{\mu\tau}|$ and $|h_{e\mu}^\dagger h_{e\tau}|$, although the couplings $h_{\mu\tau}$ and $h_{e\mu}$ could contribute to  $(g-2)_\mu$ by themselves.

The dominant SM background is from the process  $e^+ e^- \to W^+ W^- \to e^- \tau^+ \bar{\nu}_e \nu_\tau$, which is expected to be small, if we require the two charged leptons to be back-to-back and their reconstructed energy $E_{\ell} \simeq \sqrt{s}/2$.
Assuming 10 signal events as above, the prospects of the coupling $|h_{ee}^\dagger h_{e\tau}|$ are shown in the left panel of Fig.~\ref{fig:p2}. At the resonance $m_{H} \simeq \sqrt{s}$, the production cross section can be greatly enhanced by $m_{H}^2 / \Gamma_{H}^2$. To be specific, we have set the width $\Gamma_{H} = 10$ (30) GeV at $\sqrt{s} = 240$ GeV (1 TeV). For $m_H>\sqrt{s}$, the production rate diminishes rapidly as $H$ becomes heavier. An off-shell $H$ could however be probed up to a few-TeV range, as shown in Fig.~\ref{fig:p2}, and ILC is expected to be more promising than CEPC in this mass range, as a result of higher $\sqrt{s}$.

The process $e^+ e^- \to \mu^\pm \tau^\mp$ could proceed via both $s$ and $t$-channels, which depend on different couplings, namely $|h_{ee}^\dagger h_{\mu\tau}|$ and $|h_{e\mu}^\dagger h_{e\tau}|$, and are constrained respectively by the rare decays $\tau^- \to \mu^- e^+ e^-$ and $\tau^- \to \mu^+ e^- e^-$.
Analogous to the $e\tau$ case above, a broad range of $m_{H}$ and $|h_{ee}^\dagger h_{\mu\tau}|$ could be probed in the $s$-channel, especially in vicinity of the resonance, as shown by the bottom left panel of Fig.~\ref{fig:p2}. In the $t$-channel, the cross sections are comparatively smaller, and the detectable regions are much narrower, as shown by the bottom right panel of Fig.~\ref{fig:p2}. One can see that orders of magnitude of the couplings can be probed in the off-shell channels at future lepton colliders, i.e. with couplings $|h^\dagger h|$ from $\sim \, 10^{-4}$ up to ${\cal O} (1)$ for a mass range from $\sim$ 100 GeV to few TeV.

\section{LFV induced by doubly-charged scalar}
\label{sec:dcs}

In addition to the neutral scalar case, LFV could also be induced by the presence of a doubly-charged scalar, with the model-independent Yukawa couplings
\begin{eqnarray}
\label{eqn:Yukawa2}
{\cal L}_Y \ = \
f_{\alpha\beta} H^{++} \bar{\ell}_\alpha^c \ell_\beta \, +{\rm H.c.}\, ,
\end{eqnarray}
with $\alpha$, $\beta$ the flavor indices and superscript $c$ denoting charge conjugate. The doubly-charged scalar $H^{\pm\pm}$ can be either left-handed or right-handed, i.e. coupling either to the left-handed or right-handed charged leptons. We take the doubly-charged scalar $H^{\pm\pm}$ in Eq.~(\ref{eqn:Yukawa2}) to be right-handed, unless otherwise specified. The main results in this section hold also for left-handed $H^{\pm\pm}$.

At high-energy colliders, the ``smoking-gun'' signal of $H^{\pm\pm}$ are same-sign charged lepton pairs from the decay  $H^{\pm\pm} \to \ell_\alpha^\pm \ell_\beta^\pm$, mediated by the interaction in Eq.~(\ref{eqn:Yukawa2}). The current dilepton limits at the Large Hadron Collider (LHC) exclude the right-handed (left-handed) $H^{\pm\pm}$ with mass up to roughly 650 GeV (800 GeV)~\cite{ATLAS:2017iqw, CMS:2017pet}, depending on the specific lepton flavors. The final states involving only the $e$ and $\mu$ flavors are the most stringent~\cite{ATLAS:2017iqw} and those with the $\tau$ flavor are much weaker~\cite{CMS:2017pet}, as a result of the poor reconstruction efficiency of the $\tau$ lepton at hadron colliders.

There are also direct searches of doubly-charged scalars at LEP in the single~\cite{Abbiendi:2003pr} and pair~\cite{Achard:2003mv} production modes. However, limited by the center-of-mass energy, these constraints are very weak. An off-shell $H^{\pm\pm}$ in the $t$-channel could mediate the Bhabha scattering $e^+ e^- \to e^+ e^-$ and interfere with the SM diagrams. This alters both the total cross section and the differential distributions~\cite{Abbiendi:2003pr, Achard:2003mv}. By Fierz transformation, the doubly-charged scalar contributes to the effective four-fermion contact  interaction
\begin{eqnarray}
\frac{1}{\Lambda_{\rm eff}^2} (\bar{e}_R \gamma_\mu e_R) (\bar{f}_R \gamma^\mu f_R) \,,
\label{eq:eff}
\end{eqnarray}
where $\Lambda_{\rm eff} \sim M_{\pm\pm}/|f_{e\ell}|$ corresponds to the effective cutoff scale related to the doubly-charged scalar mass and the Yukawa couplings, and is constrained by the $e^+ e^- \to \ell^+ \ell^-$ (with $\ell\ell = ee,\, \mu\mu,\, \tau\tau$) data~\cite{Abdallah:2005ph}. It turns out that the cutoff scale $\Lambda_{\rm eff}$ has been excluded up to 7.6 TeV by the LEP data.

For $\alpha = \beta =e$, the effective interaction in Eq.~\eqref{eq:eff} would also induce an additional contribution to the M{\o}ller scattering $e^- e^- \to e^- e^-$ and can be constrained by the upcoming MOLLER experiment~\cite{Benesch:2014bas}, which could probe the effective scale $\Lambda_{\rm eff} \simeq 5.3$ TeV, slightly stronger than the current limit from LEP $e^+ e^-  \to e^+ e^-$ data~\cite{Dev:2018sel}. Other low-energy LFV constraints on the couplings $f_{\alpha\beta}$ include the rare decays of charged leptons like $\ell_\alpha \to \ell_\beta \gamma$ and $\ell_\alpha \to \ell_\beta \ell_\gamma \ell_\delta$, the anomalous magnetic moments of electron and muon, the LEP $e^+ e^- \to \ell^+ \ell^-$ data (with $\ell = e,\, \mu,\, \tau$)~\cite{Abdallah:2005ph}, and muonium-anti-muonium oscillation, which are all highly suppressed in the SM. The calculation details can be found in Ref.~\cite{Dev:2018upe} and references therein.

The doubly-charged scalar $H^{\pm\pm}$ can be pair-produced at lepton colliders through the Drell-Yan process via the gauge interactions with the SM photon and $Z$ boson. Drell-Yan production is, however, not a useful probe of  the Yukawa interactions $f_{\alpha\beta}$ unless $H^{\pm\pm}$ is long-lived, because for Drell-Yan pair production of $H^{\pm\pm}$, both larger and smaller Yukawa couplings $f_{\alpha\beta}$ give rise to the same line-shape over the background, and thus, we can not determine the absolute values of $f_{\alpha\beta}$ from the flavor-dependent branching fractions ${\rm BR} (H^{\pm\pm} \to \ell_\alpha^\pm \ell_\beta^\pm)$, as long as the leptonic decays dominate. Only when the Yukawa couplings enter the production of $H^{\pm\pm}$, could we measure these couplings in the (LFV) processes. All the relevant production processes can be categorized into two groups: i) the on-shell production including both the pair and single production modes and ii) the off-shell production of $H^{\pm\pm}$.

\subsection{On-shell LFV}
\label{sec:production}

\begin{figure*}[t!]
  \centering
  \includegraphics[width=0.4\textwidth]{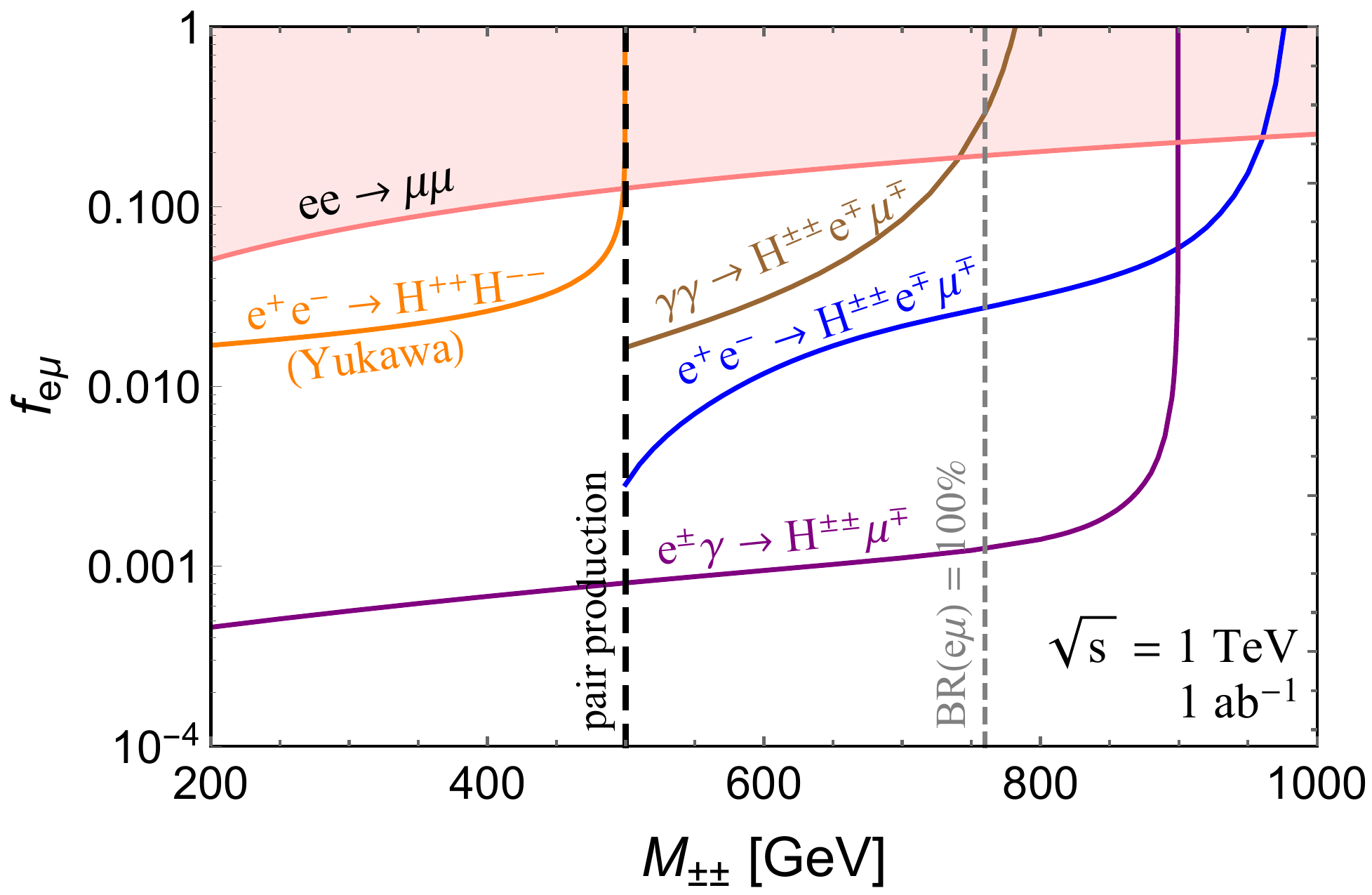} \\
  \includegraphics[width=0.4\textwidth]{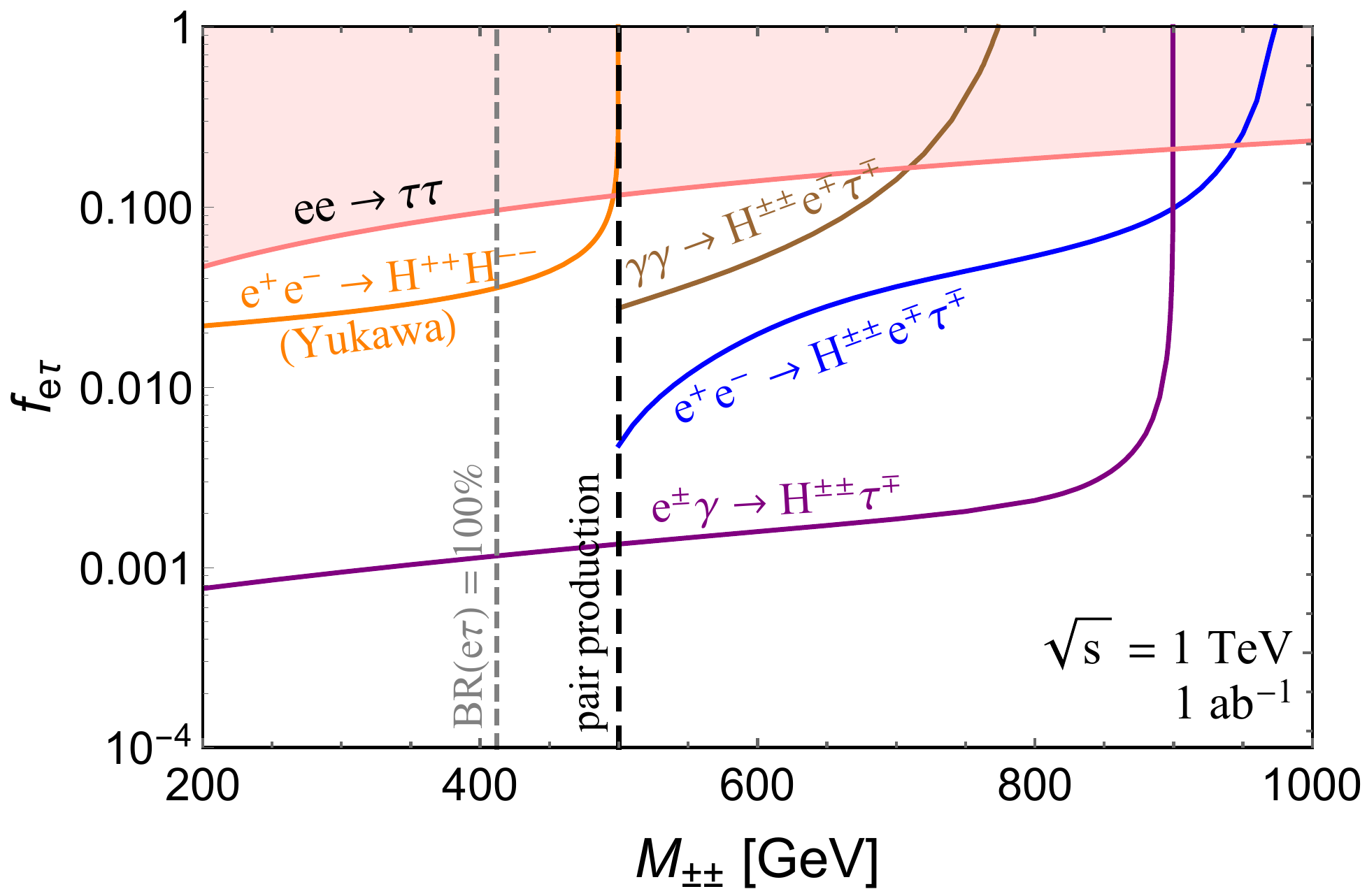}
  \includegraphics[width=0.4\textwidth]{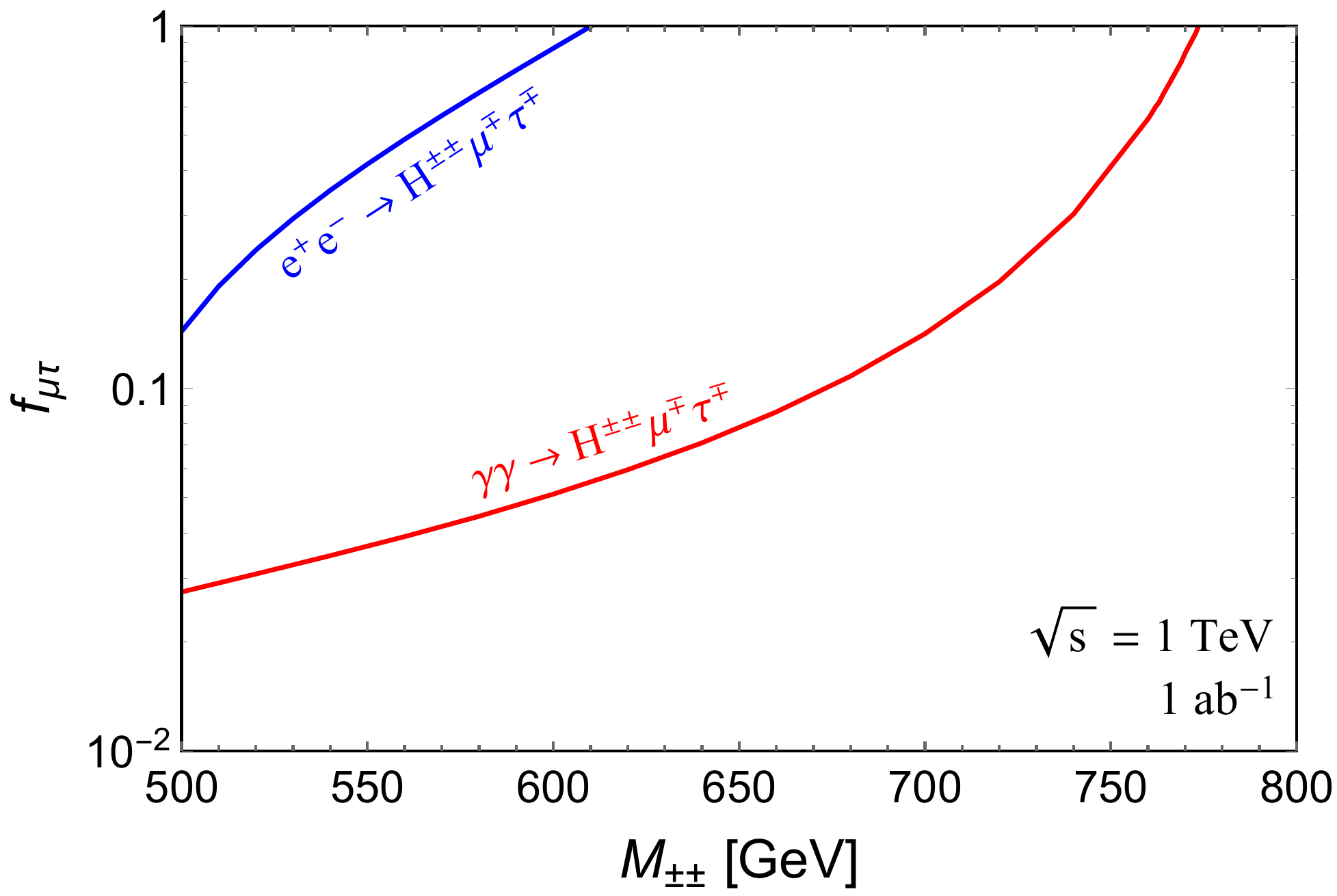}
  \caption{Prospects of the doubly-charged scalar $H^{\pm\pm}$ at ILC $1$ TeV and an integrated luminosity of 1 ab$^{-1}$.  The top and bottom left panels show both the prospects in the Yukawa pair (orange) and single production modes of the $e^+ e^-$ (blue), $e\gamma$ (purple) and $\gamma\gamma$ (brown) processes, for respectively the couplings $f_{e\mu}$ and $f_{e\tau}$, as functions of the doubly-charged scalar mass. The pink shaded regions are excluded by the LEP $ee \to \ell\ell$ data~\cite{Abdallah:2005ph}. The prospects for the Yukawa coupling $f_{\mu\tau}$ are collected in the lower right panel, in both the $e^+ e^-$ and $\gamma\gamma$ processes. The vertical dashed gray lines indicate the current same-sign dilepton limits on the doubly-charged scalar mass from LHC~\cite{ATLAS:2017iqw, CMS:2017pet}, assuming a ${\rm BR} (H^{\pm\pm} \to \ell_\alpha^{\pm} \ell_\beta^{\pm}) = 100\%$.}
  \label{fig:p3}
\end{figure*}

Besides the Drell-Yan process,\footnote{The doubly-charged scalar can also be pair-produced from the fusion of laser photons $\gamma \gamma \to H^{++} H^{--}$, induced by both the trilinear and quartic gauge-scalar interactions~\cite{Chakrabarti:1998qy}.} the doubly-charged scalar $H^{\pm\pm}$ can be pair-produced at lepton colliders through the Yukawa interactions $f_{e\ell}$ to the SM charged fermions~\cite{Kuze:2002vb},
\begin{eqnarray}
\label{eqn:dcs:pair}
e^+ e^- \to H^{++} H^{--} \quad \text{(through Yukawa couplings $f_{e\ell}$)}
\end{eqnarray}
with $\ell$ covering all three flavors $e$, $\mu$ and $\tau$.
Given the Yukawa couplings $f_{\alpha\beta}$ to the SM charged fermions, the doubly-charged scalar can be singly produced from~\cite{Barenboim:1996pt, Kuze:2002vb, Lusignoli:1989tr, Yue:2007kv, Dev:2018upe}
\begin{eqnarray}
\label{eqn:dcs:single1}
e^+ e^-,\gamma\gamma \to H^{\pm\pm} \ell_\alpha^\mp \ell_\beta^\mp
\end{eqnarray}
and~\cite{Barenboim:1996pt, Godfrey:2001xb, Godfrey:2002wp, Rizzo:1981xx, Yue:2007ym}
\begin{eqnarray}
\label{eqn:dcs:single2}
e^\pm \gamma \to H^{\pm\pm} \ell_\alpha^\mp \,,
\end{eqnarray}
with $\gamma$ the laser photon beam as in the neutral scalar case. With the same cuts as in Section~\ref{sec:neutral}, we have estimated the prospects of all the LFV couplings $|f_{e\mu,\, e\tau,\, \mu\tau}|$ at ILC 1 TeV with the integrated luminosity of 1 ab$^{-1}$, as functions the doubly-charged mass $M_{\pm\pm}$, in all the pair and single production modes in Eqs.~(\ref{eqn:dcs:pair}) to (\ref{eqn:dcs:single2}), which are collected in Fig.~\ref{fig:p3}. For simplicity, we have assumed that $H^{\pm\pm}$ decays predominately back to the same-sign charged leptons, and the electron and positron beams are both unpolarized.
There is a lower cut of 500 GeV on the single production mode in the $e^+ e^-$ and $\gamma\gamma$ single production processes, as below $\sqrt{s}/2 \simeq 500$ GeV,  the processes $e^+ e^-,\, \gamma\gamma \to H^{\pm\pm} \ell_\alpha^\mp \ell_\beta^\mp$ are dominated by the pair production with $\ell_\alpha^\mp \ell_\beta^\mp$ from the decay of an on-shell $H^{\mp\mp}$. The limits on the LFV couplings $f_{\alpha\beta}$ are mainly from the LEP $ee \to \ell\ell$ data, as a $t$-channel $H^{\pm\pm}$ could mediate the processes $ee \to \mu\mu,\,\tau\tau$ with the LFV couplings $f_{e\mu,\,e\tau}$, which exclude the shaded pink regions in Fig.~\ref{fig:p3}. The limits from electron and muon $g-2$ are highly suppressed by the charged lepton masses and are not shown in these plots. One should note that the contributions from doubly-charged
scalar loops to the  magnetic dipole moment  are always negative~\cite{Lindner:2016bgg}, so the muon $g-2$ anomaly
can not be explained by the doubly-charged scalar $H^{\pm\pm}$. The vertical dashed gray lines indicate the current same-sign dilepton limits on the doubly-charged scalar mass from LHC~\cite{ATLAS:2017iqw, CMS:2017pet}, assuming the dilepton branching of ${\rm BR} (H^{\pm\pm} \to \ell_\alpha^{\pm} \ell_\beta^{\pm}) = 100\%$. More details can be found in Ref.~\cite{Dev:2018upe}. For a smaller dilepton branching ratio, the LHC limits on doubly-charged scalar mass will be weaker~\cite{Dev:2018kpa}.

\subsection{Off-shell LFV}

\begin{figure*}[t!]
  \centering
  \includegraphics[width=0.4\textwidth]{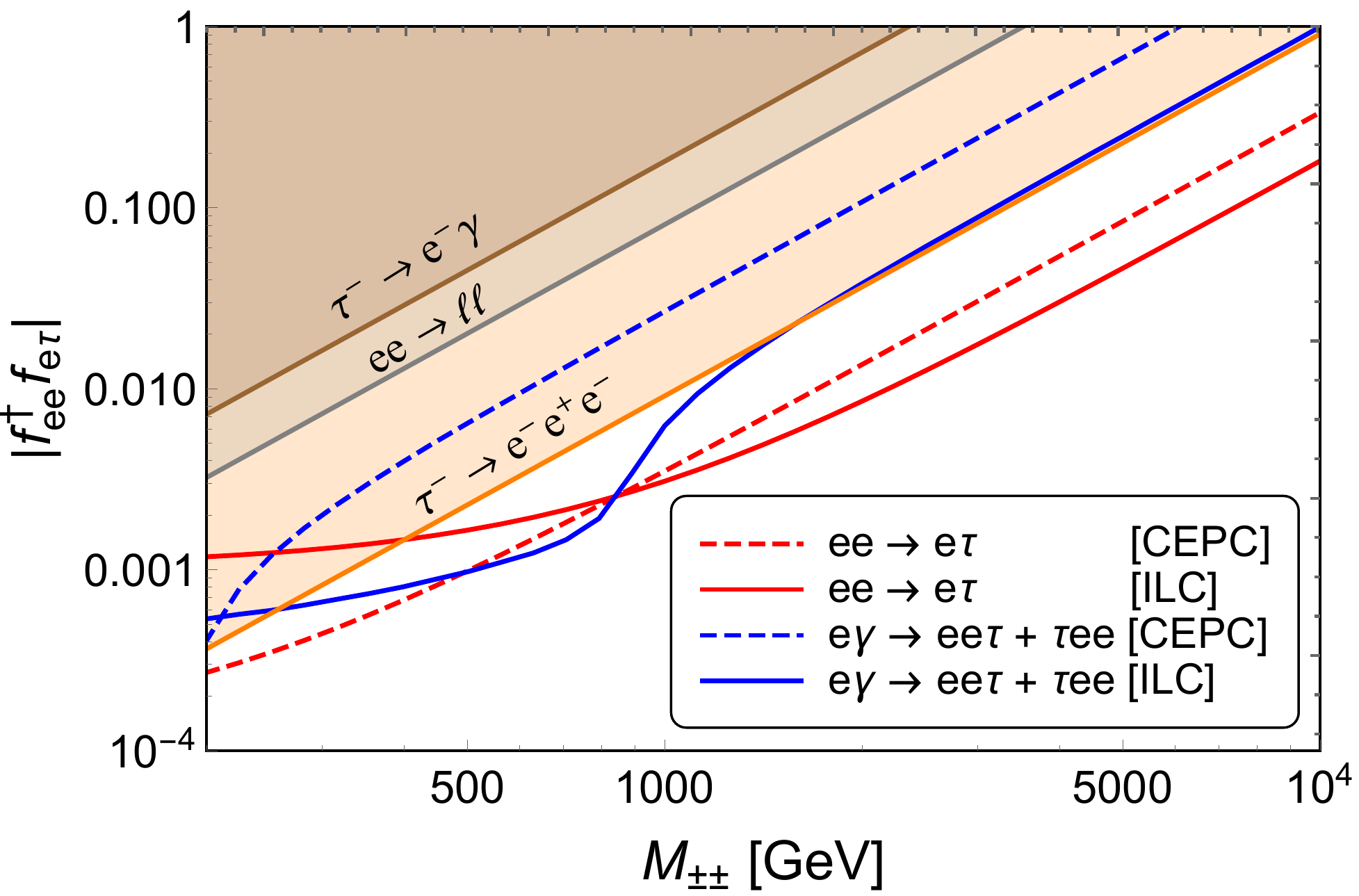}
  \includegraphics[width=0.4\textwidth]{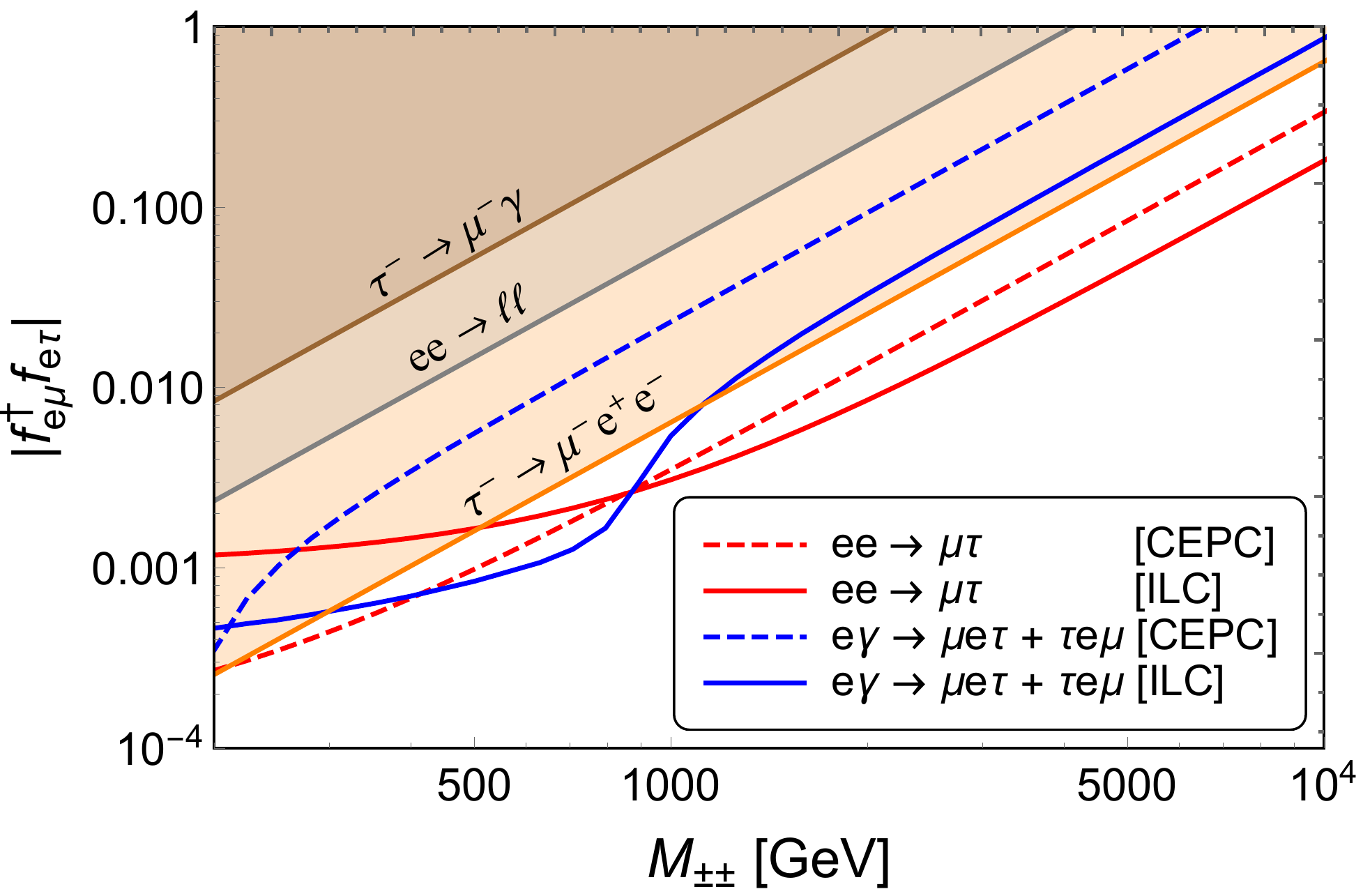}
  \caption{Prospects of the Yukawa couplings $|f_{ee}^\dagger f_{e\tau}|$ and $|f_{e\mu}^\dagger f_{e\tau}|$ for the doubly-charged scalar $H^{\pm\pm}$ production via the $ee \to \ell_\alpha \ell_\beta$ (red) and $e\gamma \to \ell_\alpha \ell_\beta \ell_\gamma$ (blue) processes, at CEPC with $\sqrt{s} = 240$ GeV and an integrated luminosity of 5 ab$^{-1}$ (dashed) and ILC with $\sqrt{s} = 1$ TeV and luminosity of 1 ab$^{-1}$ (solid). The shaded regions are excluded by the rare tau decays $\tau \to \ell_\alpha \gamma,~\ell_\alpha \ell_\beta \ell_\gamma$ and the LEP $ee \to \ell\ell$ data.}
  \label{fig:p4}
\end{figure*}

With a $t$-channel doubly-charged scalar $H^{\pm\pm}$, we can have the off-shell LFV processes~\cite{Godfrey:2001xb, Godfrey:2002wp, Rizzo:1981xx, Dev:2018upe}
\begin{eqnarray}
e^+ e^- \to \ell_\alpha^\pm \ell_\beta^\mp \quad \text{and} \quad
e^\pm \gamma \to \ell_\alpha^\mp \ell_\beta^\pm \ell_\gamma^\pm
\end{eqnarray}
which depend quadratically on the couplings $|f^\dagger f|$.  As in the neutral scalar case, the limit from $\mu \to eee$ is so stringent that it has precluded the doubly-charged scalar-mediated signal $ee \to e\mu$ at future lepton colliders. The processes $e^+ e^- \to e^\pm \tau^\mp$ ($e^+ e^- \to \mu^\pm \tau^\mp$) and $e^\pm \gamma \to e^\mp e^\pm \tau^\pm + \tau^\mp e^\pm e^\pm$ ($e^\pm \gamma \to \mu^\mp e^\pm \tau^\pm + \tau^\mp e^\pm \mu^\pm$) involve the $\tau$ lepton and depend on the coupling $|f_{ee}^\dagger f_{e\tau}|$ ($|f_{e\mu}^\dagger f_{e\tau}|$), and  their prospects at CEPC 240 GeV and ILC 1 TeV are shown in the left (right) panel of Fig.~\ref{fig:p4}. Suppressed by the three-body phase space, the sensitivities in the $e\gamma$ processes are comparatively weaker than those from $e^+ e^-$ processes.
The most important constraints on the couplings $|f_{ee}^\dagger f_{e\tau}|$ and $|f_{e\mu}^\dagger f_{e\tau}|$ are from the LFV decays $\ell_\alpha \to \ell_\beta \ell_\gamma \ell_\delta$, $\ell_\alpha \to \ell_\beta \gamma$, and the LEP $e e \to \ell \ell$ data~\cite{Abdallah:2005ph}, and these are depicted as the shaded regions in Fig.~\ref{fig:p4}~\cite{Dev:2018upe}. It is also possible to probe the LFV couplings
\begin{eqnarray}
|f_{ee}^\dagger f_{\mu\tau}|, \quad
|f_{e\mu}^\dagger f_{\mu\mu}|, \quad
|f_{e\mu}^\dagger f_{\mu\tau}|, \quad
|f_{e\mu}^\dagger f_{\tau\tau}|, \quad
|f_{e\tau}^\dagger f_{\mu\mu}|, \quad
|f_{e\tau}^\dagger f_{\mu\tau}|, \quad
|f_{e\tau}^\dagger f_{\tau\tau}|
\end{eqnarray}
in the $e\gamma$ process, however, as a result of the small production cross section, the probable parameter space is very limited for these couplings at both CEPC and ILC.

One can see in Figs.~\ref{fig:p3} and \ref{fig:p4} that in both on-shell and off-shell production modes of the doubly-charged scalar $H^{\pm\pm}$, a large parameter space of the mass $M_{\pm\pm}$ and the LFV couplings $f_{\alpha\beta}$ can be probed at ILC 1 TeV (and CEPC 240 GeV). With a luminosity of order of ab$^{-1}$,  the couplings can be probed up to the order of $10^{-3}$ in the on-shell mode, and the effective cutoff scale $\Lambda_{} \simeq M_{\pm\pm}/|f|$ could go up to few 10 TeV in the off-shell processes.

\section{Conclusion}
\label{sec:conclusion}

We have shown that a neutral scalar $H$ and doubly-charged scalar $H^{\pm\pm}$, which are both well-motivated in a large class of new physics scenarios, can be probed at future lepton colliders via their LFV couplings to the SM charged lepton sector. We present a model-independent analysis of how far the LFV coupling strengths and the scalar masses can be probed beyond the existing limits from the low-energy sector. In particular, we find that the full mass and coupling range of the neutral scalar, that can explain the muon $g-2$ anomaly, can be tested in the future lepton colliders. This is largely complementary to the searches of LFV in the low-energy experiments and  at hadron colliders.

\section*{Acknowledgments}

Y.Z.\ would like to thank the Center for High Energy Physics, Peking University for generous hospitality where the paper was finished. The work of B.D. and Y.Z. is supported by the US Department of Energy under Grant No. DE-SC0017987. The work of R.N.M was supported by the US National Science Foundation Grant No. PHY1620074.



\end{document}